\documentstyle [12pt,a4]{article}
\def\p{\partial}
\def\l{\lambda}
\begin{document}

\thispagestyle{empty}
\begin{flushright}
MPI/95-136 \\
HUB-IEP/95-32 \\
hep-th/9512130 \\
December 1995
\end{flushright}

\vspace{2cm}

\begin{center}
{\Large \bf Conformal Symmetries \\
\vspace{2mm}
of the Self-Dual Yang-Mills Equations}
\vspace{1cm}

{\large A.D.~Popov}\footnote {On leave of absence from Bogoliubov Laboratory
of Theoretical Physics, JINR, Dubna, Russia},\\
\vspace{2mm}
{\em Max-Planck-Institut f\"{u}r Mathematik,\\
Gottfried-Claren-Str. 26, 53225 Bonn, Germany}

\vspace{4mm}

{\large C.R.~Preitschopf}\\
\vspace{2mm}
{\em Institut f\"{u}r Physik, Humboldt-Universit\"{a}t zu Berlin,\\
Invalidenstr. 110, 10115 Berlin, Germany}\\
\end{center}

\vspace{2cm}

\begin{center}
{\large \bf Abstract}
\end{center}

\begin{quote}
We describe an infinite-dimensional Kac-Moody-Virasoro
algebra of new hidden symmetries for the self-dual Yang-Mills
equations related to conformal transformations of the
4-dimensional base space.
\end{quote}

\newpage

\section{Introduction}

The self-dual Yang-Mills (SDYM) equations in the space $R^4$
with the metric of signature $(+ + + +)$
or $(+ +  - -)$ are the famous example of the nonlinear integrable
equations in four dimensions. These equations are invariant under
the group of gauge transformations and the group of conformal
 transformations
 of the space  $R^4$, both of which are the ``obvious symmetries".
It is well-known by now that the SDYM equations in $R^4$ have an
infinite-dimensional
algebra of ``hidden symmetries" [1-5]. For the Yang-Mills (YM) potentials with
values in a Lie algebra $\cal G$ these symmetries are an affine
extension of the
Lie algebra $\cal G$ of global gauge transformation to the Kac-Moody
algebra ${\cal G}\otimes C(\l , \l^{-1})$. It is also well-known
that for integrable models in two dimensions the algebra of hidden
symmetries
includes Virasoro-like generators (for a recent discussion and references
see [6]). We shall show that the SDYM equations also have Virasoro-like and
new Kac-Moody symmetries, which generate new solutions from any old one.

New Kac-Moody-Virasoro symmetries may be interesting for applications
because
\begin{itemize}
\item  they give
new arguments in support of the old idea that the SDYM theory
is a generalization of the $d=2$ conformal theories to dimension $d=4$;
\item the SDYM equations are known to arise in $N=2$ supersymmetric open
string theory and Kac-Moody-Virasoro symmetries may underlie the
cancellation of
almost all amplitudes in the theory of $N=2$ self-dual strings;
\item the study of symmetries is important for understanding non-pertubative
properties and quantization of Yang-Mills theories and $N=2$ self-dual strings;
\item the SDYM equations are ``master" integrable equations since a lot of
integrable equations in $1\le d\le 3$ can be obtained from them by suitable
reductions (for a recent discussion and references see [7]). In particular,
the matrix Ernst-type equations appearing in the study of
T- and S-duality symmetries of string theory may be obtained by reduction
of the SDYM equations. Their symmetries in turn may also be obtained
by an appropriate reduction of symmetries of the SDYM equations.
\end{itemize}

Clearly this list of reasons is not complete and can be extended.

\section{Definitions and notation}

We shall consider the Euclidean space $R^{4,0}$ with the
metric $g_{\mu\nu}=diag (+1,$  $ +1,$ $ +1,$ $ +1)$, where $\mu ,
\nu ,...=1,...,4$.
Let us denote by $A_\mu$ the potentials of the YM fields
$F_{\mu\nu}=\p_\mu A_\nu - \p_\nu A_\mu + [A_\mu , A_\nu ]$,
with $\p_\mu := \p / \p x^\mu$. The fields $A_\mu$ and $F_{\mu\nu}$
take values in a Lie algebra $\cal G$. For simplicity one may think that
${\cal G}\  =\  sl(n, C)$.

The SDYM equations have the form
$$
\frac{1}{2} \varepsilon _{\mu\nu\rho\sigma}F^{\rho\sigma}=F_{\mu\nu},
\eqno(1)
$$
where $\varepsilon _{\mu\nu\rho\sigma}$ is the completely antisymmetric
tensor in $R^{4,0}$ ( $\varepsilon _{1234}=1$). In $R^{4,0}$ we introduce
complex coordinates
$y=x^1+i x^2, z=x^3 -
ix^4, {\bar y}=x^1 - i x^2,
{\bar z} = x^3+i x^4$
and set
$A_y=\frac{1}{2}(A_1-i A_2), A_z=\frac{1}{2}(A_3 +
iA_4), A_{\bar y}=\frac{1}{2}(A_1 + i A_2),
A_{\bar z} = \frac{1}{2}(A_3-i A_4)$.
The SDYM equations read then
$$
F_{yz}=0, \quad F_{\bar y\bar z}=0,
\eqno(2a)
$$
 $$F_{y\bar y} + F_{z\bar z}=0.
\eqno(2b)$$
These equations can be obtained as compatibility conditions of the following
linear system of equations [8]:
$$
[\p_{\bar y} + A_{\bar y} - \l (\p_z + A_z)] \Psi (x, \l) =0,
\eqno(3a)
$$
$$[\p_{\bar z} + A_{\bar z} + \l (\p_y + A_y)] \Psi (x, \l) =0,
\eqno(3b)
$$
where $\Psi \in G$ is a group-valued function depending on the coordinates
$x^\mu$ of $R^{4,0}$ and a complex parameter $\l\in CP^1$.
In fact, $\Psi$ is defined on the twistor space ${\cal Z}=R^{4,0}\times CP^1$
for the space $R^{4,0}$, and eqs. (3) are equivalent to the holomorphicity of
the matrix-function $\Psi$ (Ward theorem [9]).

Equation (2a) implies that the gauge potentials can be written in the
form
$$
A_y = h^{-1}\p_y h, \ A_z= h^{-1}\p_z h, \ A_{\bar y}=
{\tilde h}^{-1}\p_{\bar y} {\tilde h}, \ A_{\bar z}=
{\tilde h}^{-1}\p_{\bar z} {\tilde h},
\eqno(4)
$$
where $h$ and $\tilde h$ are some group-valued functions on
$R^{4,0}$. One may perform the following gauge transformation:
$$
A_{\bar y} \rightarrow B_{\bar y} = {\tilde h} A_{\bar y} {\tilde h}^{-1}+
{\tilde h} \p_{\bar y} {\tilde h}^{-1}=0,\
A_{\bar z} \rightarrow B_{\bar z} = {\tilde h} A_{\bar z} {\tilde h}^{-1}
+{\tilde h} \p_{\bar z} {\tilde h}^{-1}=0,\
\eqno(5a)
$$
$$
A_{y} \rightarrow B_{y} = {\tilde h} A_{ y} {\tilde h}^{-1}+
{\tilde h} \p_{y} {\tilde h}^{-1}=g^{-1}\p_y g,\
A_{z} \rightarrow B_{z} = {\tilde h} A_{ z} {\tilde h}^{-1}+
{\tilde h} \p_{z} {\tilde h}^{-1}=g^{-1}\p_z g,
\eqno(5b)
$$
where $g:= h{\tilde h}^{-1}$, and thus fix the gauge
$B_{\bar y}=B_{\bar z}=0\ $  [10,1-5].
Then eqs. (2) are replaced by the matrix equation
$$
\p_{\bar y} B_y + \p_{\bar z} B_z=
\p_{\bar y}(g^{-1}\p_y g) + \p_{\bar z} (g^{-1}\p_zg)=0.
\eqno(6)
$$
It is also possible to perform the gauge transformation
$$
A_{\bar y}\rightarrow  h A_{\bar y} h^{-1}+
h \p_{\bar y} {h}^{-1}= g\p_{\bar y} g^{-1},\
A_{\bar z}\rightarrow  {h} A_{\bar z} { h}^{-1}
+{h} \p_{\bar z} { h}^{-1}=g \p_{\bar z} g^{-1},\
\eqno(7a)
$$
$$
A_{y}\rightarrow {h} A_{ y} {h}^{-1}+{h} \p_{y} {h}^{-1}=0,\
A_{z} \rightarrow {h} A_{ z} {h}^{-1}+{h} \p_{z} {h}^{-1}=0,
\eqno(7b)
$$
then eq.(2) gets converted into the equation
$$ \p_y(g\p_{\bar y} g^{-1}) + \p_z (g\p_{\bar z}g^{-1}) =
-g \left[ \p_{\bar y}(g^{-1}\p_y g) +
\p_{\bar z} (g^{-1}\p_zg) \right] g^{-1}=0.
\eqno(8)
$$

Let $C$ denote a contour in the $\l$-plane about the origin,
$C_+$ denote $C$ and the inside of $C$, and $C_-$ denotes $C$ and
the outside of $C$ ($C=C_+\cap C_-$). Then there exist two matrix
functions
$\Psi_{\pm}(x, \l )$ holomorphic and nonsingular on $C_{\pm}$,
each satisfying eqs. (3) (see e.g. [5]). From the linear system (3)
it is easy to see that
$$
h=\Psi_-^{-1}(\l =\infty ),\quad {\tilde h}=\Psi_+^{-1}(\l =0 ).
\eqno(9)
$$
Therefore, it is obvious that eqs. (6) are the compatibility conditions
of the linear system
$$
\p_{\bar y}\eta - \l (\p_z +B_z)\eta =0,
\eqno(10a)
$$
$$
\p_{\bar z}\eta + \l (\p_y +B_y)\eta =0,
\eqno(10b)
$$
obtained from (3)  for $\Psi_+$ by performing the gauge transformation
$\Psi_+(x,\l )\rightarrow \eta (x,\l )$ = $\Psi_+^{-1}(x,0)\Psi_+(x,\l )$ =
${\tilde h}(x)\Psi_+(x,\l ), \l \in C_+$.
Analogously, eqs. (8) are compatibility conditions for the linear
system
$$
\frac{1}{\l}(\p_{\bar y} +g\p_{\bar y}g^{-1})\hat\eta  -
\p_z\hat\eta =0,
\eqno(11a)
$$
$$
\frac{1}{\l}(\p_{\bar z} + g\p_{\bar z}g^{-1})\hat\eta
+\p_y\hat\eta =0,
\eqno(11b)
$$
where
$\hat\eta (\l )=\Psi_-^{-1}(\infty)\Psi_-(\l )$ is well defined for
$\l\in C_-$.

In the following, we shall consider the space ${\cal Z}_+=R^{4,0}\times
C_+\subset {\cal Z}=R^{4,0}\times CP^1$, matrix-valued function $\eta$ on
${\cal Z}_+$, the
linear system (10) and the Wess-Zumino-Novikov-Witten-type eq.(6).
In this short paper we  describe new hidden symmetries of
eqs. (6) omitting direct computations and writing out only the final
formulas.

\section{Space-time symmetries}

Let $A_\mu$ be a solution of the SDYM equations (1). Then $\delta : A_\mu
\rightarrow \delta A_\mu$ is called
an infinitesimal symmetry transformation if $\delta A_\mu$ satisfies
the linearized form of eqs.(1). It is well-known that the group of
conformal transformations is the maximal group
of transformations of the
space $R^{4,0}$ under which the SDYM equations (1)
are invariant. This group is locally isomorphic to the group
$SO(5,1)$.

Let us introduce the self-dual $\eta^a_{\mu\nu}$ and the anti-self-dual
$\bar\eta^a_{\mu\nu}$ 't Hooft tensor,
$$
\eta^a_{\mu\nu}=\{\epsilon^a_{bc}, \mu =b, \nu =c;\ \delta^a_\mu , \nu =4;
-\delta^a_\nu , \mu =4 \},
\eqno(12a)
$$
$$
\bar\eta^a_{\mu\nu}=\{\epsilon^a_{bc}, \mu =b, \nu =c;\ -\delta^a_\mu , \nu =4;
\delta^a_\nu , \mu =4 \},
\eqno(12b)
$$
where $a,b,... = 1,2,3$ and $\epsilon^a_{bc}$ are the structure constants of
the group $SO(3)$. Then the generators of the group $SO(5,1)$ can be realized
in terms of the following vector fields on $R^{4,0}$,
$$
X_a = \delta_{ab}\eta^b_{\mu\nu} x_\mu\p_\nu ,\
Y_a = \delta_{ab}\bar\eta^b_{\mu\nu} x_\mu\p_\nu ,\  P_\mu =\p_\mu ,
$$
$$
K_\mu =\frac{1}{2}x_\sigma x_\sigma \p_\mu - x_\mu D ,
\quad D=x_\sigma\p_\sigma ,
\eqno(13)
$$
where $\{X_a\}$ and $\{Y_a\}$ generate two commuting $SO(3)$-subgroups in
$SO(4)$, $P_\mu$ are the generators of translations, $K_\mu$  are the
generators of special conformal transformations and $D$ is
the generator of dilatations.

Infinitesimal transformations of the YM potentials $A_\mu$ under the
action of the conformal group $SO(5,1)$ are given by
$$
\delta_N A_\mu ={\cal L}_N A_\mu := N^\nu A_{\mu ,\nu} + A_\nu N^\nu _{,\mu}
\quad,
\eqno(14)
$$
where $N=N^\nu \p_\nu$ is any generator of the conformal group,
and ${\cal L}_N$ is the Lie
derivative along the vector field $N$. It is not hard to show that for each
$N\in so(5,1)$ the transformation (14) constitutes a symmetry of the
SDYM equations (1).

The group-valued functions $\Psi_\pm$ satisfying the linear equations (3)
are defined on the subspaces ${\cal Z}_\pm =R^{4,0}\times C_\pm$ of
the twistor space ${\cal Z}$.
Therefore, we have to define the action of $SO(5,1)$ on ${\cal Z}_\pm $
preserving the linear system (3). The lifted vector fields $\tilde N$
on ${\cal Z}_+$, which form the generators of $SO(5,1)$, are given by [7],
$$
\tilde X_a = X_a, \  \tilde Y_a = Y_a + 2 Z_a, \   \tilde P_\mu = P_\mu ,
$$
$$
\tilde K_\mu = K_\mu + \bar\eta^a_{\sigma\mu}x_\sigma Z_a ,\quad \tilde D=D,
\eqno(15)
$$
with the following expression for the generators $Z_a$ of the $SO(3)$
rotations on $C_+\subset CP^1$:
$$
Z_1 =\frac{i}{2} (\l^2-1)\p_\l -\frac{i}{2} (\bar\l^2-1)\p_{\bar\l },\
Z_2 =\frac{1}{2} (\l^2+1)\p_\l +\frac{1}{2} (\bar\l^2+1)\p_{\bar\l },\
Z_3= i\l\p_\l - i\bar\l\p_{\bar\l }.
\eqno(16)
$$
Here $\bar\l$ is the complex conjugate to $\l\in C_+$ and $\p_\l := \p
/\p\l $.
For ${\cal Z}_-$ we have analogous formulas. Using identities for the
't Hooft
tensors [11], it can be easily shown that $[X_a, X_b]= -2\epsilon^c_{ab} X_c$,
$[X_a, Y_b]=0, [\tilde Y_a, \tilde Y_b]= -2\epsilon^c_{ab}\tilde Y_c$ and so
on.
Now one can define the infinitesimal transformation of
the group-valued function
$\Psi_+$ under the action of the conformal group [7]:
$$
\delta_{\tilde N }\Psi_+ ={\cal L}_{\tilde N }\Psi_+ := \tilde N \Psi_+,
\eqno(17)
$$
where $\tilde N$ is any of the generators defined
in (15). It is not hard to show that the
linear system (3) is invariant under the transformations (14), (17).

Obviously the gauge $B_{\bar y} = B_{\bar z} =0$ is not invariant
under conformal transformations, because in general
$({\cal L}_NB_{\bar y})\mid _{B_{\bar y}=0}\not= 0, ({\cal L}_NB_{\bar z})
\mid _{B_{\bar z}=0}\not= 0$. However, as noticed by Pohlmeyer [1],
conformal invariance can be restored by compensating gauge transformations.
We shall write out the explicit formulas for these compensating gauge
transformations. First, it can be shown that
$$
\delta_{X_a}B_{\bar y}={\cal L}_{X_a} B_{\bar y}=B_y X^y_{a,\bar y} +
B_z X^z_{a,\bar y}=0, \quad \delta_{X_a}B_{\bar z}=0,
$$
$$
\delta_{P_\mu}B_{\bar y}=\delta_{P_\mu}B_{\bar z}=0, \
\delta_{D}B_{\bar y}=\delta_{D}B_{\bar z}=0, \
\delta_{Y_3}B_{\bar y}=\delta_{Y_3}B_{\bar z}=0,
\eqno(18)$$
and for $N\in \{Y_1, Y_2, K_\mu\}$ we have $\delta_{N}B_{\bar y}\not=0$ and
$\delta_{N}B_{\bar z}\not=0.$ For example,
$$
\delta_{Y_2}B_{\bar y}= -B_z,\
\delta_{Y_2}B_{\bar z}= B_y,
\eqno(19a)
$$
$$
\delta_{K_y}B_{\bar y}=-zB_z,\ \delta_{K_y}B_{\bar z}=zB_y,
\eqno(19b)
$$
where $B_y$ and $B_z$ are defined in (5).

As a second step, for an arbitrary generator $\tilde N$ of (15) let us
introduce the Lie algebra-valued function
$\psi_{\tilde N}(x,\l )$ on ${\cal Z}_+$,
$$
\psi_{\tilde N}(x,\l )=(\tilde N(\l ) \eta (\l ))\eta^{-1}(\l ) \in {\cal G},
\eqno(20)
$$
and consider the function
$\varphi_N=\psi^0_{\tilde N}(x)=$ $\psi_{\tilde N}(x, \l =0)$,
the infinitesimal gauge transformation
$\p_\mu\varphi_N + [B_\mu , \varphi _N]$ of the YM potentials
generated by $\varphi_N(x)$, and the transformation
$$
\delta^0_{\tilde N} B_y = \delta_N B_y + \p_y\varphi_N + [B_y, \varphi_N],\
\delta^0_{\tilde N} B_z = \delta_N B_z + \p_z\varphi_N + [B_z, \varphi_N],
\eqno(21a)
$$
which is a combination of the conformal transformations
$\delta_N$ and a gauge transformation. By direct calculation
one may verify that
$$
\delta^0_{\tilde N} B_{\bar y} = \delta_N B_{\bar y} +
\p_{\bar y}\varphi_N =0,\
 \delta^0_{\tilde N} B_{\bar z} = \delta_N B_{\bar z} +
\p_{\bar z}\varphi_N=0,
\eqno(21b)
$$
for any generator $\tilde N$ of the conformal group. The identity
$\eta (x,\l =0)=1$ implies that $\varphi_N=0$ and
$\delta^0_{\tilde N}=\delta_N$
for $N\in\{P_\mu , X_a, D, Y_3\}$ and $\varphi_N\not=0$
if $N\in \{Y_1, Y_2, K_\mu \}$. Using eqs. (14) and (21), it is easy
to show that
$$
[\delta^0_{\tilde M}, \delta^0_{\tilde N}] B_y=
\delta^0_{[\tilde M,\tilde N]} B_y,\
[\delta^0_{\tilde M}, \delta^0_{\tilde N}] B_z=
\delta^0_{[\tilde M,\tilde N]} B_z,\
[\delta^0_{\tilde M}, \delta^0_{\tilde N}] B_{\bar y}= 0,
[\delta^0_{\tilde M}, \delta^0_{\tilde N}] B_{\bar z}= 0,
\eqno(22)
$$
i.e. the transformations $\delta^0_{\tilde N}$ define
an action of the conformal  algebra $so(5,1)$ on the YM potentials which
preserves the gauge $B_{\bar y}=B_{\bar z}=0$.

The action of $\delta^0_{\tilde N}$ on the
group-valued function $\eta (x, \l )$ has the form
$$
\delta^0_{\tilde N}\eta (\l ) = \delta_{\tilde N}\eta (\l ) -\varphi_N(x)
\eta (\l )=
\tilde N(\l )\eta (\l ) -\varphi_N(x) \eta (\l ),
\eqno(23)
$$
where $\delta_{\tilde N}^0$ is precisely the combination of a conformal
and a compensating gauge transformation. It may be shown that the
linear system (10) is invariant under the transformations (21), (23).
Thus, fixing a complex structure on the space $R^{4,0}$ and fixing the
gauge (5) does not destroy the conformal invariance of the SDYM equations
and the linear system for them.

\section{Hidden symmetries}

In Sect.3 we assigned
to each generator $\tilde N$ of the conformal group $SO(5,1)$
a function $\psi_{\tilde N}(x, \l )$
on ${\cal Z}_+$ with values in the Lie algebra $\cal G$
(see (20)). Using eqs. (10) on $\eta (x,\l )$, one can
verify that the function $\psi_{\tilde N}$ satisfies the
equations
$$
\p_{\bar y}\psi_{\tilde N}(\l ) +{\cal L}_N B_{\bar y}
- \l (\p_z\psi_{\tilde N}(\l )+[B_z, \psi_{\tilde N}(\l )]+
{\cal L}_N B_z)=0,
\eqno(24a)
$$
$$
\p_{\bar z}\psi_{\tilde N}(\l ) +{\cal L}_N B_{\bar z}
+\l (\p_y\psi_{\tilde N}(\l )+[B_y, \psi_{\tilde N}(\l )]+
{\cal L}_N B_y)=0,
\eqno(24b)
$$
where ${\cal L}_N$ is the Lie derivative along the vector field
 $N$ (see Sect.3).

Let us recall that the group-valued function $\eta (x,\l )$ is
holomorphic and nonsingular for $\l \in C_+$. Hence
$\psi_{\tilde N}$ can be expanded in powers of $\l $:
$$
\psi_{\tilde N} (x, \l )=\sum^\infty_{n=0}\l^n
\psi_{\tilde N}^n(x),
\eqno(25)
$$
where the coefficients $\psi_{\tilde N}^n$ depend only on
 $x^{\mu}\in R^{4,0}$ and are conserved nonlocal charges (cf. [1-4]).
After substituting (25) into (24) we
obtain the following recurrence relations:
$$
{\cal L}_N B_{\bar y} + \p_{\bar y}\psi_{\tilde N}^0 =0,\
{\cal L}_N B_{\bar z} + \p_{\bar z}\psi_{\tilde N}^0 =0,
\eqno(26a)
$$
$$
\p_{\bar y}\psi_{\tilde N}^1 - \p_z\psi_{\tilde N}^0 -
[B_z, \psi_{\tilde N}^0 ]-{\cal L}_N B_z=0,\
 \p_{\bar z}\psi_{\tilde N}^1 + \p_y\psi_{\tilde N}^0 +
[B_y, \psi_{\tilde N}^0 ]+{\cal L}_N B_y=0,
\eqno(26b)
$$
$$
\p_{\bar y}\psi_{\tilde N}^{n+1} - \p_z\psi_{\tilde N}^n -
[B_z, \psi_{\tilde N}^n ]=0,\
 \p_{\bar z}\psi_{\tilde N}^{n+1} + \p_y\psi_{\tilde N}^n +
[B_y, \psi_{\tilde N}^n ]=0,\ n\ge 1 .
\eqno(26c)
$$
The starting point (26a) is true by construction of $\delta^0_{\tilde N}$:
$$
\delta^0_{\tilde N}B_{\bar y}=0,\ \delta^0_{\tilde N}B_{\bar z}=0.
\eqno(27a)
$$
For $B_y$ and $B_z$ we obtain
$$
\delta^0_{\tilde N}B_y:= \p_y\psi_{\tilde N}^0 +[B_y,
\psi_{\tilde N}^0]+{\cal L}_NB_y
=-\p_{\bar z}\psi_{\tilde N}^1,
\eqno(27b)
$$
$$
\delta^0_{\tilde N}B_z:=\p_z\psi_{\tilde N}^0 +
[B_z,\psi_{\tilde N}^0]+{\cal L}_NB_z
=\p_{\bar y}\psi_{\tilde N}^1.
\eqno(27c)
$$
Using eqs. (27) it is trivial to deduce the
invariance of the SDYM equations (6)
under conformal transformations:
$$
\p_{\bar y}(\delta^0_{\tilde N}B_y) +\p_{\bar z}(\delta^0_{\tilde N}B_z)=
-\p_{\bar y}\p_{\bar z}
\psi_{\tilde N}^1 +\p_{\bar z}\p_{\bar y}\psi_{\tilde N}^1 =0.
\eqno(28)
$$
Eqs. (27) are straightforward to generalize to an infinite
number of infinitesimal transformations $\delta^n_{\tilde N}$
with $n\ge 0$:
$$
\delta^n_{\tilde N}B_{\bar y}:= 0,\quad \delta^n_{\tilde N}
B_{\bar z}:= 0,\quad n\ge 1,
\eqno(29a)
$$
$$
\delta^n_{\tilde N}B_{ y}:= \p_y\psi_{\tilde N}^n +[B_y,\psi_{\tilde N}^n]=
-\p_{\bar z}\psi_{\tilde N}^{n+1},
\  n\ge 1,
\eqno(29b)
$$
$$
\delta^n_{\tilde N}B_{z}:= \p_z\psi_{\tilde N}^n +[B_z,\psi_{\tilde N}^n]=
\p_{\bar y}\psi_{\tilde N}^{n+1},
\  n\ge 1.
\eqno(29c)
$$
Introducing $\delta_{\tilde N}(\zeta )=
\sum^\infty_{n=0}\zeta^n\delta^n_{\tilde N},\
\zeta\in C_+$,
we obtain a one-parameter family of infinitesimal transformations
$$
\delta_{\tilde N}(\zeta )B_{\bar y}= 0,\quad
\delta_{\tilde N}(\zeta )B_{\bar z}= 0,
\eqno(30a)
$$
$$
\delta_{\tilde N}(\zeta )B_{ y}:= \p_y\psi_{\tilde N}(\zeta )
+[B_y,\psi_{\tilde N}(\zeta )] +
{\cal L}_N B_y= -\frac{1}{\zeta}(\p_{\bar z}\psi_{\tilde N}(\zeta )
+ {\cal L}_N B_{\bar z}),
\eqno(30b)
$$
$$
\delta_{\tilde N}(\zeta )B_{z}:= \p_z\psi_{\tilde N}(\zeta ) +
[B_z,\psi_{\tilde N}(\zeta )]+
{\cal L}_N B_z=\frac{1}{\zeta}(\p_{\bar y}
\psi_{\tilde N}(\zeta )+{\cal L}_N B_{\bar y}).
\eqno(30c)
$$
Clearly $\delta_{\tilde N}(\zeta )$ generates a
symmetry of the SDYM equations
(6), since if $B_y$ and $B_z$ satisfy eqs. (6) then
$$
\p_{\bar y}(\delta_{\tilde N}(\zeta )B_y) +\p_{\bar z}(\delta _{\tilde N}
(\zeta ) B_z)=
-\frac{1}{\zeta}\{\p_{\bar y }({\cal L}_N B_{\bar z})-\p_{\bar z}({\cal L}_N
B_{\bar y})\} =0,
\eqno(31)
$$
for any generator $\tilde N$ of the conformal group.
For example, for $Y_2$ from (19a) we have
$$
\p_{\bar y}(\delta_{\tilde Y_2}(\zeta )B_y) +
\p_{\bar z}(\delta _{\tilde Y_2}(\zeta ) B_z)=
-\frac{1}{\zeta}\{\p_{\bar y}B_{ y}+\p_{\bar z}B_{z}\} =0.
$$

To find the infinitesimal transformation
$\eta (\l )\rightarrow \delta_{\tilde N}(\zeta )\eta (\l )$
corresponding to the transformations (30) let us consider the variation of
eqs. (10). We obtain
$$
\p_{\bar y}\chi_{\tilde N}(\l , \zeta ) - \l (\p_z\chi_{\tilde N}(\l , \zeta )
+[B_z,\chi_{\tilde N}(\l , \zeta )]) -\l\delta_{\tilde N}(\zeta )B_z =0,
\eqno(32a)
$$
$$
\p_{\bar z}\chi_{\tilde N}(\l , \zeta ) + \l (\p_y\chi_{\tilde N}(\l , \zeta )
+[B_y,\chi_{\tilde N}(\l , \zeta )]) + \l\delta_{\tilde N}(\zeta )B_y  =0,
\eqno(32b)
$$
where $\chi_{\tilde N}(\l , \zeta )=
\{\delta_{\tilde N}(\zeta )\eta(\l )\}\eta^{-1}(\l )$.
One can verify that the function $\chi_{\tilde N}(\l , \zeta )$
satisfies eqs. (32)
(cf. [1-6]):
$$
\chi_{\tilde N}(\l , \zeta )=
\frac{\l }{\l -\zeta }\{\psi _{\tilde N}(\l ) - \psi_{\tilde N}(\zeta )\}
\Rightarrow
$$
$$
\delta_{\tilde N}(\zeta )\eta (\l )= \frac{\l }{\l -\zeta }
\{\psi _{\tilde N}(\l ) - \psi_{\tilde N}(\zeta )\}
\eta (\l )\ .
\eqno(33)
$$
For $\zeta =0$ formula (33) coincides with eq. (23).
Thus, we succeeded to assign
to each generator $\tilde N$ of the conformal group
a one-parameter family $\delta_{\tilde N}(\zeta),\
\zeta\in C_+$, of infinitesimal transformations of
solutions of the SDYM equations and of the solution $\eta$
of the associated linear system.
For each $\tilde N\in so(5,1)$ these
transformations are new ``hidden symmetries" of the SDYM equations.

\section{Algebraic structure of the symmetries}

In Sect.4 we described the infinitesimal symmetry transformations,
the exponentiation of which acts on the set $\cal M$
of solutions of the SDYM equations (6). Let us consider any
solution $\{B_{\mu }\}=\{B_y, B_z, B_{\bar y}=0, B_{\bar z}=0\}$
of these equations. Then the solutions $\delta _{\tilde N} (\zeta )B_{\mu }$
of the linearized SDYM equations describe the vector space
tangent to the manifold $\cal M$ of solutions at the point $\{B_{\mu }\}$.
So, the infinitesimal symmetries $\delta _{\tilde N}(\zeta )$
are vector fields on the manifold $\cal M$
and they define a map $B_{\mu }\rightarrow \delta _{\tilde N}
(\zeta )B_{\mu }$.

Notice that we consider $\l , \zeta \in C_+$.
We restrict our attention to only half of the symmetries.
The rest will be obtained when we focus on
${\cal Z}_-=R^{4,0}\times C_-$, the linear system (11) for $\hat \eta$ on
${\cal Z}_-$ and the function $\hat\psi_{\tilde N}(x,\zeta )$ $=
\{\tilde N(\zeta )\hat\eta (\zeta )\}\hat\eta ^{-1}(\zeta ), \zeta\in C_-$.
We recall that the
function $\hat \eta$ is holomorphic and nonsingular for $\zeta\in C_-$
and therefore $\hat\psi_{\tilde N}(\zeta )=\sum^{\infty}_{n=0}\zeta^{-n}
\hat\psi_{\tilde N}^n$. Then we can derive for  $\hat\psi_{\tilde N}$ an
equation analogous to eq.(24) and introduce
a second set of symmetry transformations  $\hat\delta _{\tilde N} (\zeta )$
similar to (30) and (33) with $\hat\psi_{\tilde N}$
replacing $\psi_{\tilde N}$.

Let us discuss now the algebraic properties of the ``on-shell"
symmetry transformations $\delta_{\tilde N}(\zeta )$ that preserve
the equations of motion. After a
lengthy computation using the formulas of Sect.3 and Sect.4,
we obtain the following expression for the commutator between
two successive infinitesimal transformations:
$$
[\delta_{\tilde M}(\l ),\delta_{\tilde N}(\zeta )]B_y =
\frac{1}{\l -\zeta }\{\l \delta_{[\tilde M,\tilde N]}(\l )-\zeta
\delta_{[\tilde M,\tilde N]}(\zeta )\}B_y-
\eqno(34a)
$$
$$
-\frac{1}{\l\zeta (\l -\zeta )^2}\left\{
\zeta^2 \tilde M^{\l}\left (\l\delta_{\tilde N}
(\l )-\zeta\delta_{\tilde N}(\zeta )\right )
+\l^2 \tilde N^{\zeta}\left(\l\delta_{\tilde M}
(\l )-\zeta\delta_{\tilde M}(\zeta )\right )\right\}B_y+
\eqno(34b)
$$
$$
+\frac{1}{\l -\zeta }\{ \tilde M^{\zeta }\p_{\zeta}
(\zeta\delta_{\tilde N}(\zeta ))+
\tilde N^{\l}\delta_{\l}(\l \delta_{\tilde M}(\l ))
\}B_y+
\eqno(34c)
$$
$$
+\{\frac{1}{\zeta}\tilde N^y_{,\bar z}\delta_{\tilde M}(\l )
-\frac{1}{\l}\tilde M^y_{,\bar z}\delta_{\tilde N}(\zeta )\}B_y -
\eqno(34d)
$$
$$
-\frac{1}{\l -\zeta }\{\tilde M^\zeta_{,\bar z}\p_{\zeta}\psi_{\tilde N}
(\zeta ) +\tilde N^{\zeta}_{,\bar z}\p_{\l}\psi_{\tilde M}(\l )\}+
\eqno(34e)
$$
$$
+
\frac{\l\tilde N^{\zeta}_{,\bar z}}{\zeta(\l -\zeta )^2}\{\psi_{\tilde M}
(\l )-\psi_{\tilde M}(\zeta )\}
+\frac{\zeta\tilde M^{\l}_{,\bar z}}{\l (\l -\zeta )^2}\{\psi_{\tilde N}
(\l )-\psi_{\tilde N}(\zeta )\}.
\eqno(34f)
$$
Here ${\tilde N}^y,{\tilde N}^{\l},...$ are the components of any generator
of $SO(5,1)$ of (15): $\tilde N = {\tilde N}^y\p_y +
{\tilde N}^{\bar y}\p_{\bar y}+{\tilde N}^z\p_z + {\tilde N}^{\bar z}
\p_{\bar z}+{\tilde N}^{\l}\p_{\l}+{\tilde N}^{\bar \l}\p_{\bar \l}.$
The commutator $[\delta_{\tilde M}(\l ), \delta_{\tilde N}(\zeta )]B_z$
looks similar to (34), except $\tilde N^z_{,\bar y},
\tilde M^z_{,\bar y}, \tilde N^{\zeta }_{,\bar y}$ and
$\tilde M^{\zeta }_{,\bar y}$ replace $\tilde N^y_{,\bar z}$ etc.
in lines (34d)-(34f).

It is obvious that because of the terms (34e) and (34f) the
commutator (34) does not close
in general. The terms (34e) and (34f) are nonzero
when $\tilde N^{\l}_{,\bar y}\ne 0$ and $\tilde N^{\l}_{,\bar z}\ne 0$,
but this holds only for the generators ${\tilde K}_{\mu}$ of special
conformal transformations. Therefore, if at least one of the vector fields
$\tilde M$ or $\tilde N$ coincides with one of the generators
$\tilde K_{\mu}$, then the commutator of two symmetries is no
longer a symmetry.

Now we consider the 8-dimensional
algebra $\cal A$ with generators $\{P_{\mu}, X_a, D\}$ and the 11-dimensional
algebra $\cal B$ with generators $\{P_{\mu}, X_a, D,\tilde Y_a\}$. Both
algebras $\cal A$ and $\cal B$ are subalgebras of $so(5,1)$.
For each generator $N$ of the algebra $\cal A\subset\cal B$ we have
$\tilde N=N$ ($\tilde P_{\mu}=P_{\mu}, \tilde X_a=X_a, \tilde D=D$),
i.e. these generators have no components along the vector fields $\p_{\l}$
and $\p_{\bar\l}$. Moreover, for all of them $\tilde N^y_{,\bar z}=
\tilde N^z_{,\bar y}=0$
and hence all terms (34b)-(34f) are zero. Thus, for $\tilde M,
\tilde N\in\cal A$ eq. (34) reduces to
$$
[\delta_M(\l ), \delta_N(\zeta )]=\frac{1}{(\l -\zeta )}\{\l\delta_{[M,N]}
(\l )-\zeta\delta_{[M,N]}(\zeta )\},
\eqno(35)
$$
which defines the analytic half of the Kac-Moody algebra
${\cal A}\otimes C(\l ,\l^{-1})$, the affine extension of the algebra
${\cal A}\in so(5,1)$.
Let us define the variations $\delta^n_N$ for all $n\ge 0$ by the contour
integral
$$
\delta_N^n=\oint_{C^\prime}\frac{d\l}{2\pi i}\l^{-n-1}\delta_N(\l ),
\eqno(36)
$$
where the contour $C^\prime$ circles once around $\l =0$. We may
choose $C^\prime=C=C_+\cap C_-$.
Using the definition (36) and the commutator (35), Cauchy's theorem
allows us to deduce the commutators between half of the generators
of the affine Lie algebra ${\cal A}\otimes C(\l , \l^{-1})$:
$$
[\delta_M^m,\delta_N^n]=\delta_{[M,N]}^{m+n},\quad m,n\ge 0.
\eqno(37)
$$

For the generators $\{\tilde Y_a\}$ of the $so(3)$-subalgebra of the algebra
$\cal B$ we have $\tilde Y^{\l}_{a,\bar y}=\tilde Y^{\l}_{a,\bar z}=0$,
${\tilde Y}^{z}_{a,\bar y}=const$,
${\tilde Y}^{y}_{a,\bar z}=const$ and thus the terms (34e) and (34f)
are zero. Instead of $\{\tilde Y_a\}$ it is convenient
to rewrite the generators as follows:
$$
\tilde Y_+:=\tilde Y_2-i\tilde Y_1=2z\p_{\bar y}-2y\p_{\bar z}+2(Z_2-iZ_1),
\eqno(38a)
$$
$$
\tilde Y_-:= \tilde Y_2+i\tilde Y_1= 2\bar z\p_y -2\bar y\p_z +2(Z_2+iZ_1),
\eqno(38b)
$$
$$
\tilde Y_0:=-i\tilde Y_3=y\p_y+z\p_z-\bar y\p_{\bar y}-\bar z\p_{\bar z}-2iZ_3,
\eqno(38c)
$$
$$
[\tilde Y_0,\tilde Y_+]=2\tilde Y_+,
\ [\tilde Y_0,\tilde Y_-]=-2\tilde Y_-,\
[\tilde Y_+,\tilde Y_-]=-4\tilde Y_0.
\eqno(38d)
$$
Using the explicit form (38) of the vector fields $\tilde Y_0, \tilde Y_{\pm}$
and eqs. (34) we obtain:
$$
[ \delta^m_{\tilde Y_0}, \delta^n_{\tilde Y_0}] = -4(m-n)
\delta^{m+n}_{\tilde Y_0},\ \delta^m_{\tilde Y_0}:=
\oint_C\frac{d\l }{2\pi i}\l^{-m-1}\delta_{\tilde Y_0}(\l ),
\eqno(39a)
$$
$$
[ \delta^m_{\tilde Y_+}, \delta^n_{\tilde Y_+}] = -4(m-n)
\delta^{m+n-1}_{\tilde Y_+},
\ \delta^m_{\tilde Y_{+}}:=
\oint_C\frac{d\l }{2\pi i}\l^{-m-1}\delta_{\tilde Y_{+}}(\l ),
\eqno(39b)
$$
$$
[ \delta^m_{\tilde Y_-}, \delta^n_{\tilde Y_-}] = -4(m-n)
\delta^{m+n+1}_{\tilde Y_-},
\ \delta^m_{\tilde Y_{-}}:=
\oint_C\frac{d\l }{2\pi i}\l^{-m-1}\delta_{\tilde Y_{-}}(\l ).
\eqno(39c)
$$

{}From (39) one can see that $\delta^m_{\tilde Y_0}$, $\delta^m_{\tilde Y_+}$
and $\delta^m_{\tilde Y_-}$
generate three different Virasoro-like subalgebras of the symmetry algebra.
Obviously these Virasoro-like subalgebras do not commute with each other.
Using eq.(34), the contour integral definitions and Cauchy's theorem one finds
$$
[\delta^m_{\tilde Y_0},\delta^n_N]=\delta^{m+n}_{[Y_0,N]}+4n\delta^{m+n}_N,
\eqno(40a)
$$
$$
[\delta^m_{\tilde Y_+},\delta^n_N]=\delta^{m+n}_{[Y_+,N]}+4n\delta^{m+n-1}_N,
\eqno(40b)
$$
$$
[\delta^m_{\tilde Y_-},\delta^n_N]=\delta^{m+n}_{[Y_-,N]}+4n\delta^{m+n+1}_N,
\eqno(40c)
$$
$$
[\delta^m_{\tilde Y_0},\delta^n_{\tilde Y_+}]=
\delta^{m+n}_{[\tilde Y_0,\tilde Y_+]}+4n\delta^{m+n}_{\tilde Y_+}-
4m\delta^{m+n-1}_{\tilde Y_0},
\eqno(40d)
$$
$$
[\delta^m_{\tilde Y_0},\delta^n_{\tilde Y_-}]=
\delta^{m+n}_{[\tilde Y_0,\tilde Y_-]}+
4n\delta^{m+n}_{\tilde Y_-}-4m\delta^{m+n+1}_{\tilde Y_0},
\eqno(40e)
$$
$$
[\delta^m_{\tilde Y_+},\delta^n_{\tilde Y_-}]=\delta^{m+n}_{[\tilde Y_+,
\tilde Y_-]}+4n\delta^{m+n-1}_{\tilde Y_-}-4m\delta^{m+n+1}_{\tilde Y_+}.
\eqno(40f)
$$
Thus, the subset of symmetries of the SDYM equations with generators
$\delta^m_{P_{\mu }}, \delta^m_{X_a}, \delta^m_D$ and
$\delta^m_{\tilde Y_a}$
forms a Kac-Moody-Virasoro algebra with commutation relations
(37), (39) and (40).

In [7] it has been shown that the linear system for the
SDYM equations will be invariant under the action of the conformal group
 only if we add the combinations of the Virasoro generators
$L_0=-\l\p_{\l}, L_1=-{\l}^2\p_{\l}$ and $L_{-1}=-\p_{\l}$ to the generators
 $Y_a$ and $K_{\mu}$ . Therefore, the appearance
of a Virasoro-like algebra as an algebra of symmetries is not surprising.
Beyond expectation the symmetry algebra contains {\it three}
different Virasoro-like subalgebras with the generators
$\delta^m_{\tilde Y_0}, \delta^m_{\tilde Y_+}$ and $\delta^m_{\tilde
Y_-}$.
In comparison with the previously known symmetries of the SDYM equations,
these new symmetries are the affine extension not of
gauge symmetries, but of space-time symmetries of the SDYM equations.

\section{Off-shell Kac-Moody and Virasoro algebras}

We will now proceed to define the ``off-shell"
action of the graded Lie algebra $so(5,1)\otimes C(\l )$ and the Virasoro
algebra on the space of YM potentials and group-valued functions
 $\eta$. The action of the Virasoro algebra does not preserve the SDYM
equations. As to the algebra $so(5,1)\otimes C(\l )$, only the action of the
subalgebra ${\cal A}\otimes C(\l )$ of this algebra preserves eq. (6).

As usual, we assign to each vector field $N$ the function
$\psi_N(x,\zeta )=\{N\eta (\zeta )\}\eta^{-1}(\zeta )$ on
${\cal Z}_+$
and consider the following transformation of the YM potentials:

$$
\delta_N(\zeta)B_y:= \p_y\psi_N(\zeta ) +[B_y, \psi_N(\zeta )]
+{\cal L}_NB_y, \quad \delta _N(\zeta)B_{\bar y}:={\cal L}_NB_{\bar y},
\eqno(41a)
$$
$$
\delta _N(\zeta)B_z:= \p_z\psi_N(\zeta ) +[B_z, \psi_N(\zeta )]
+{\cal L}_NB_z, \quad \delta _N(\zeta)B_{\bar z}:={\cal L}_NB_{\bar z}.
\eqno(41b)
$$
For $\eta (x,\l )$ we postulate the following transformation rule:
$$
\delta _N(\zeta)\eta (\l ):= \frac{\l }{\l -\zeta }
\{\psi_N(\l ) -\psi_N(\zeta )\}\eta(\l ).
\eqno(41c)
$$
Now it is not hard to show that
\setcounter{equation}{41}
\begin{eqnarray}
&[\delta_M(\l ), \delta_N(\zeta )]B_y=
	&\p_y\{\delta_N(\zeta )\psi_M(\l ) - \delta_M(\l )\psi_N(\zeta )\}+
							\nonumber\\
&& +[B_y, \delta_N(\zeta )\psi_M(\l )-\delta_M(\l )\psi_N(\zeta )]+
							\nonumber\\
&& +[\delta_N(\zeta )B_y, \psi_M(\l )] - [\delta_M(\l )B_y, \psi_N(\zeta )]+
							\nonumber\\
&& + {\cal L}_M\{\delta_N(\zeta )B_y\} -{\cal L}_N\{\delta_M(\zeta )B_y\}
\end{eqnarray}
and we have the same formula for $B_z$. From (41), (42) and
$$
\delta_N(\zeta )\psi_M(\l )=\{M(\delta _N(\zeta)\eta (\l ))\}\eta^{-1}(\l )
+\{M\eta (\l ))\}\delta _N(\zeta)\eta^{-1}(\l )
$$
it follows that
$$
[\delta_M(\l ), \delta_N(\zeta )]=\frac{1}{(\l -\zeta )}\{\l\delta_{[M,N]}
(\l )-\zeta\delta_{[M,N]}(\zeta )\},
\eqno(43)
$$
when we consider the action on $B_y, B_z, B_{\bar y}$ and $B_{\bar z}$, as
well as $\eta$.
By (41) we have for $B_{\bar y}$ or $B_{\bar z}$ simply
$$
[\delta_M(\l ), \delta_N(\zeta )]B_{\bar y}=\frac{1}{(\l -\zeta )}
\{\l\delta_{[M,N]}(\l )-\zeta\delta_{[M,N]}(\zeta )\}B_{\bar y}=
{\cal L}_{[M,N]}B_{\bar y}.
\eqno(44)
$$

Using eq. (43), it is not difficult to deduce the commutators of half of the
affine Lie algebra $\widehat{so}(5,1)$:
$$
[\delta_M^m,\delta_N^n]=\delta_{[M,N]}^{m+n}, \quad
\delta_N^n=\oint_C\frac{d\l}{2\pi i}\l^{-n-1}\delta_N(\l ),
\quad m,n\ge 0.
\eqno(45)
$$
If we consider also $\hat\eta$, $\hat\psi_N$ and generators $\hat\delta_N$,
then we obtain (45) with $-\infty\le m,n\le +\infty$, i.e. the full
affine extention $\widehat{so}(5,1)=so(5,1)\otimes C(\l ,\l^{-1})$
of the conformal algebra $so(5,1)$.

Consider now the vector field $V=\l\p_{\l}$ on $C_+$ and the
Lie algebra-valued function
$$
\psi_V(x,\l )=\{V(\l )\eta (\l )\}\eta^{-1}(\l ).
\eqno(46)
$$
Let us define for $\{B_{\mu }\}$ and $\eta$ the following transformation
 rules
$$
\delta (\zeta)B_y:= \p_y\psi_V(\zeta ) +[B_y, \psi_V(\zeta )],\quad
\delta (\zeta )B_{\bar y}:=0,
\eqno(47a)
$$
$$
\delta (\zeta)B_z:= \p_z\psi_V(\zeta ) +[B_z, \psi_V(\zeta )],\quad
\delta (\zeta )B_{\bar z}:=0,
\eqno(47b)
$$
$$
\delta (\zeta )\eta (\l ):=\frac{\l }{\l -\zeta }
\{\psi_V(\l ) -\psi_V(\zeta )\}\eta(\l ).
\eqno(47c)
$$
One can verify that
$$
[\delta (\l ), \delta (\zeta )]B_y=\p_y\{\delta (\zeta )\psi_V(\l ) -
\delta (\l )\psi_V(\zeta )\}+$$
$$ +[B_y, \delta (\zeta )\psi_V(\l )-
\delta (\l )\psi_V(\zeta )]+[\delta (\zeta )B_y, \psi_V(\l )] -
[\delta (\l )B_y, \psi_V(\zeta )],
\eqno(48)
$$
and similarly for $B_z$. Using (47), (48) and
$$
\delta (\zeta )\psi_V(\l ) =\frac{\l }{\l -\zeta }
\{V(\l )\psi_V(\l )+ [\psi_V(\l ), \psi_V(\zeta )]\}-
\frac{\l \zeta }{(\l -\zeta )^2}\{\psi_V(\l ) -\psi_V(\zeta )\}
$$
we obtain the commutator
$$
[\delta (\l ), \delta (\zeta )]=\frac{1}{(\l -\zeta )}
\{\l^2\p_{\l }\delta (\l ) +\zeta^2\p_{\zeta}\delta(\zeta )\}-
\frac{2\l\zeta}{(\l -\zeta )^2}\{\delta (\l )-\delta (\zeta )\}
\eqno(49)
$$
when we consider the action on $\eta$ or on any component of $B_{\mu}$.

Let us define
$$
L^n=-\frac{1}{2}\delta^n=-\frac{1}{2}\oint_C\frac{d\l}{2\pi
 i}\l^{-n-1}\delta (\l ).
\eqno(50)
$$
Now, using (49) and (50), it is not hard to show that we obtain half
of the Witt algebra:
$$
[L^m, L^n]=(m-n)L^{m+n},\quad m,n\ge 0.
\eqno(51)
$$
We shall obtain the full centerless Virasoro algebra, i.e. the Witt
algebra, if we consider
$\hat\eta(\zeta )$ with $\zeta\in C_-$ and extend all the calculations
appropriately.

Finally, we write out the  formula
for the commutator between the generators $L^m$ of the Virasoro algebra
and the generators $\delta_N^n$ of the Kac-Moody algebra
$\widehat{so}(5,1)$:
$$
[\delta (\l ), \delta_N(\zeta )]=\frac{\zeta^2}{(\l -\zeta )}
\p_{\zeta }\delta_N(\zeta ) - \frac{\l\zeta}{(\l -\zeta )^2}
\{\delta_N(\l )- \delta_N(\zeta )\}\ \Rightarrow
\eqno(52a)
$$
$$
[L^m, \delta_N^n]=-n\delta_N^{m+n}.
\eqno(52b)
$$
Thus we have defined the off-shell action of the Kac-Moody algebra
$\widehat{so}(5,1)$ and of the Virasoro algebra on the space of
YM potentials and on the group-valued functions $\eta (\l )$.

\newpage

{\Large \bf Acknowledgements}

\vspace{5mm}

A.D.P. is grateful to Yu.I.Manin and W.Nahm for discussions,
while C.R.P. would like to acknowledge conversations with
I.Ya.Aref'eva and C.Devchand. For its hospitality A.D.P. thanks the
Institut f\"{u}r Physik, Humboldt-Universit\"{a}t zu Berlin,
where part of this work was done, and the DAAD for support.

\vspace{5mm}

{\Large \bf References}
\begin{enumerate}
\item K.Pohlmeyer, Commun.Math.Phys. {\bf 72} (1980) 37.
\item L.-L.Chau, M.-L.Ge and Y.-S.Wu, Phys.Rev. {\bf D25} (1982)
1086;\\
 L.-L.Chau and Y.-S.Wu, Phys.Rev. {\bf D26} (1982) 3541;\\
L.-L.Chau, M.-L.Ge, A.Sinha and Y.-S.Wu, Phys.Lett. {\bf B121} (1983) 391;\\
L.-L.Chau, Lect. Notes Phys. Vol.189 (1983) 111.
\item K.Ueno and Y.Nakamura, Phys.Lett. {\bf B109} (1982) 273.
\item L.Dolan, Phys.Lett. {\bf B113} (1982) 387; Phys.Rep. {\bf 109} (1984) 3.
\item L.Crane, Commun.Math.Phys. {\bf 110} (1987) 391.
\item J.H.Schwarz, Nucl.Phys. {\bf B447} (1995) 137; Nucl.Phys.
{\bf B454} (1995)  427.
\item M.Legar\'{e} and A.D.Popov, Phys.Lett. {\bf A198} (1995) 195;\\
T.A.Ivanova and A.D.Popov, Theor.Math.Phys. {\bf 102} (1995) 280;\\
T.A.Ivanova and A.D.Popov, Phys.Lett. {\bf A205} (1995) 158.
\item  R.S.Ward, Phys.Lett. {\bf A61} (1977) 81;\\
A.A.Belavin and V.E.Zakharov,
Phys.Lett. {\bf B73} (1978) 53.
\item R.S.Ward and R.O.Wells Jr., {\it Twistor geometry and field theory},\\
Cambridge Univ.Press, Cambridge, 1990.
\item C.N.Yang, Phys.Rev.Lett. {\bf 38} (1977) 1377.
\item M.K.Prasad, Physica {\bf D1} (1980) 167.
\end{enumerate}
\end {document}